\documentstyle[12pt]{article}
\begin{document}
\topmargin=-1.0cm
\evensidemargin=0cm
\oddsidemargin=0cm
\newcommand{\BQ}{\begin{equation}}
\newcommand{\EQ}{\end{equation}}
\newcommand{\BQA}{\begin{eqnarray}}
\newcommand{\EQA}{\end{eqnarray}}
\newcommand{\half}{\frac{1}{2}}
\newcommand{\NN}{\nonumber \\}
\newcommand{\E}{{\rm e}}
\newcommand{\del}{\partial}
\renewcommand{\thefootnote}{\fnsymbol{footnote}}
\newcommand{\zm}[1]{\stackrel{\circ} {#1} }
\newcommand{\nzm}[1]{\widetilde {#1} }
\newcommand{\llangle}{\langle \langle}
\newcommand{\rrangle}{\rangle \rangle}
\date{}
\baselineskip=0.6cm
\renewcommand{\appendix}{\renewcommand{\thesection}
{\Alph{section}}\renewcommand{\theequation}{\Alph{section}.\arabic{equation}}
\setcounter{equation}{0}\setcounter{section}{0}}
\begin{titlepage}
\begin{flushright}
{KOBE-TH-99-01\\ IFUP-TH 6/99} 
\end{flushright}
\vspace{3cm}
\begin{center}
{\LARGE Spontaneously Broken Translational
Invariance of}\\
\vspace{5mm}
{\LARGE Compactified Space}\\
\vskip1.0truein
{\large Makoto Sakamoto}$^{(a)}$,
\footnote{E-mail: {\tt sakamoto@oct.phys.kobe-u.ac.jp}}
{\large Motoi Tachibana}$^{(b)}$ and 
\footnote{E-mail: {\tt tatibana@oct.phys.kobe-u.ac.jp}}
{\large Kazunori Takenaga}$^{(c)}$ 
\footnote{E-mail: {\tt takenaga@ibmth.df.unipi.it}} 
\vskip0.2truein
\centerline{$^{(a)}$ {\it Department of Physics,
Kobe University, Rokkodai, Nada, Kobe 657-8501, Japan}}
\vspace*{2mm}
\centerline{$^{(b)}$ {\it Graduate School of Science and Technology,
Kobe University, Rokkodai, Nada, Kobe 657-8501, Japan }}
\vspace*{2mm}
\centerline{$^{(c)}$ {\it I.N.F.N, Sezione di Pisa, Via Buonarroti, 2 Ed. B, 
56127 Pisa, Italy}}
\end{center}
\vskip0.5truein 
\centerline{\bf Abstract} 
\vskip0.13truein
We propose a mechanism to break the translational invariance of compactified 
space spontaneously. 
As a simple model, we study a real $\phi^4$ model compactified on
$M^{D-1}\otimes S^1$ in detail, where we impose a nontrivial boundary 
condition on $\phi$ for the $S^1$-direction.
It is shown that the translational invariance 
for the $S^1$-direction is spontaneously
broken when the radius $R$ of $S^1$ becomes larger than a critical
radius $R^*$ and also that the model behaves like a $\phi^4$ model on a
single kink background for $R \rightarrow \infty$. It is pointed out
that spontaneous breakdown of translational invariance is accompanied by 
that of some global symmetries, in general, in our mechanism.
\end{titlepage}
\newpage
\baselineskip 20 pt
\pagestyle{plain}
\vskip0.2truein
\vskip0.2truein
Spacetime translational invariance can be broken spontaneously
if any local operators acquire spacetime-dependent vacuum expectation
values. This situation, however, seems implausible because a configuration
which minimizes a potential is, in general,  independent of
spacetime coordinates and further because the spacetime-dependent
vacuum configuration
would produce nonzero kinetic energy,
so that energetically such the configuration would be
unfavorable. One might expect that even if some of space dimensions
are compactified,
the translational invariance of the compactified space
(if exists) could not be broken spontaneously.
\par
The purpose of this paper is to propose a mechanism to break 
the translational
invariance of compactified spaces spontaneously. To illustrate our
mechanism, let us consider a real $\phi^4$ model in $D$ dimensions
\BQ
S = \int d^D x \left\{- \half \del_M\phi \del^M\phi - V(\phi)\right\},
\label{action}
\EQ
where the index $M$ runs from 0 to $D-1$ and
\BQ
V(\phi) 
= \frac{\lambda}{8}\left(\phi^2 - 2\frac{\mu^2}{\lambda}\right)^2.
\label{potential}
\EQ
We should note that the action has a $Z_2$ symmetry
\BQ
\phi \rightarrow -\phi.
\EQ 
It turns out that the existence of global symmetries is crucial to
our mechanism and that the above $Z_2$ symmetry plays an important
role in this model. One may conclude that the ground state would
be given by $\phi = \pm{\sqrt{\frac{2}{\lambda}}\mu}$, at which the
scalar potential $V(\phi)$ takes the minimum value and the $Z_2$
symmetry is broken. This is, however, a hasty conclusion, as we 
will see below.
\par
Let us suppose that one of the space coordinates, say, $y \equiv
x^{D-1}$ is compactified on a circle $S^1$ whose radius is $R$.
Since $S^1$ is multiply-connected and the action has the $Z_2$
symmetry, we can impose the following nontrivial boundary condition
associated with the $Z_2$ symmetry for the field $\phi$:
\BQ
\phi(x^{\mu},y+2\pi R) = -\phi(x^{\mu}, y),
\label{bc}
\EQ
where $x^{\mu}$ denote 
the coordinates of the uncompactified spacetime. 
Thanks to the
$Z_2$ symmetry, the action (\ref{action}) is still single-valued even
with the boundary condition (\ref{bc}). An important consequence of 
the nontrivial boundary condition (\ref{bc}) is that any vacuum
expectation value of $\phi(x^{\mu}, y)$ cannot be a ($y$-independent)
nonzero constant, which is inconsistent with the boundary
condition (\ref{bc}). In other words, any nonzero vacuum expectation
value of $\phi(x^{\mu}, y)$ should have the $y$ dependence
in order to satisfy the boundary condition (\ref{bc}), i.e.
\BQ
\langle \phi(x^{\mu}, y)\rangle \neq 0 \quad \longrightarrow 
\quad \frac{\del}{\del y}\langle \phi(x^{\mu}, y) \rangle \neq 0. 
\label{y-dep}
\EQ
It immediately follows that if the vacuum is translationally
invariant the vacuum expectation value of $\phi(x^{\mu}, y)$ has to
vanish, or conversely that if $\phi(x^{\mu}, y)$ acquires a nonzero
vacuum expectation value, which implies the $y$ dependence of
$\langle \phi(x^{\mu}, y) \rangle$, 
then the translational invariance for the $S^1$-direction is 
spontaneously broken.
\par
In order to find a vacuum configuration, one might try to minimize
the potential $V(\phi)$. This is, however, wrong in the present
model. To find a vacuum configuration, we should take account of
kinetic terms in addition to potential terms since the translational
invariance could be broken and then the vacuum configuration might
be coordinate-dependent. Since the translational invariance of the  
(uncompactified) $(D-1)$-dimensional Minkowski spacetime is expected 
to be unbroken, finding a vacuum configuration of the model may be
equivalent to solving a minimization problem of the following
functional
\footnote{The ${\cal E}[\phi]$ may be regarded as a potential
from a viewpoint of the $(D-1)$-dimensional Minkowski spacetime.}:
\BQ
{\cal E}[\phi,R] \equiv \int^{2\pi R}_{0}dy
\left\{ \half(\del_y \phi)^2 + V(\phi) \right\}.
\label{energy}
\EQ
In this paper, our analysis will be restricted to the tree level.
\par
In the following, we shall ignore the $x^{\mu}$ dependence in $\phi$
since we are interested in a vacuum configuration, for which
the
translational invariance of the $(D-1)$-dimensional Minkowski
spacetime is assumed to be unbroken. It should be emphasized that
$\phi(y)$ cannot be an arbitrary function but has to obey the
(antiperiodic) boundary condition
\BQ
\phi(y+2\pi R) = -\phi(y).
\label{bc2}
\EQ
If 
the translational invariance for the $S^1$-direction 
is unbroken, the vacuum expectation 
value of $\phi$ has to vanish and then the functional ${\cal E}[\phi,R]$
becomes
\BQ
{\cal E}[\phi=0,R] = \frac{\pi R \mu^4}{\lambda}.
\label{energy2}
\EQ
If there exists any configuration of $\phi(y)$ such that
\BQ
{\cal E}[\phi,R] <
{\cal E}[\phi=0,R] = \frac{\pi R \mu^4}{\lambda},
\label{energy3}
\EQ
then the configuration $\phi = 0$ is no longer a vacuum configuration
and hence 
the translational invariance for the $S^1$-direction 
has to be broken, as 
discussed previously. 
Before minimizing the functional ${\cal E}[\phi,R]$,
we would like to make a comment about solutions to the field equation
for $\phi(y)$
\BQA
0 &=& \frac{\delta {\cal E}[\phi,R]}{\delta \phi(y)} \NN 
  &=& -\frac{d^2\phi(y)}{dy^2}+\frac{\lambda}{2}\phi(y)
\left( (\phi(y))^2- 2\frac{\mu^2}{\lambda} \right ).
\label{fieldeq}
\EQA
We first note that $\phi=0$ is a trivial solution to eq.(\ref{fieldeq}),
while another trivial (would-be) solutions $\phi= 
\pm{\sqrt{\frac{2}{\lambda}}\mu}$ have to be excluded due to the 
boundary condition (\ref{bc2}). Using the above field equation to
eliminate the term $\half(\del_y\phi(y))^2$ in eq.(\ref{energy}),
we find
\BQA
{\cal E}[\phi,R]\Big|_{\frac{\delta {\cal E}}{\delta \phi}=0}
&=& \frac{\pi R \mu^4}{\lambda}- \int^{2\pi R}_{0}dy
\frac{\lambda}{8}(\phi(y))^4 \NN
&\leq& {\cal E}[\phi=0,R].
\label{energy4}
\EQA
Since the equality on the last line holds only when $\phi = 0$,
we have thus arrived at an important conclusion that if there
appear nontrivial solutions $\phi$ to the field equation (\ref{fieldeq}),
then ${\cal E}[\phi,R]$ is lower than ${\cal E}[0,R]$ so that
the translational invariance 
for the $S^1$-direction is broken spontaneously
with the $Z_2$ symmetry breaking.
\par
Let us now proceed to find a vacuum configuration, which minimizes
the functional ${\cal E}[\phi,R]$. To this end, we shall first
construct whole solutions to the field equation (\ref{fieldeq})
with the boundary condition (\ref{bc2}), which are candidates of
a vacuum configuration. We shall then determine which configuration
gives the lowest value of ${\cal E}[\phi,R]$ (if there exist
several solutions). The field equation (\ref{fieldeq}) has been studied
before in a quite different context \cite{Manton,Liang}, though
the boundary condition has been imposed to be periodic but not
antiperiodic. It turns out that most of the results given in
ref.\cite{Manton,Liang} are useful for our purposes and that
the nontrivial solutions to our problem will be given by
\BQ
\phi(y) = \frac{2k\omega}{\sqrt{\lambda}}{\rm sn}(\omega (y-y_0),k),
\label{sol}
\EQ
where
\BQ
\omega \equiv \frac{\mu}{\sqrt{1+k^2}}.
\label{omega}
\EQ
Here, ${\rm sn}(u,k)$ is the Jacobi elliptic function whose period is 
$4K(k)$, where $K(k)$ denotes the complete elliptic function of the 
first kind. Since 
the integration constant $y_0$ in eq.(\ref{sol}),
which in fact reflects the translational invariance
of the equation of motion, is irrelevant,
we shall set $y_0$ to be equal to zero in the following analysis.
The antiperiodic boundary condition (\ref{bc2}) requires that the
parameter $k$ ($0 \leq k < 1$) and the radius $R$ should be related
mutually through
\BQ
R = (2n-1)\frac{K(k)}{\pi \omega}
\label{radius}
\EQ
for some positive integer $n$. (For the periodic boundary condition,
$2n-1$ in eq.(\ref{radius}) should be replaced by $2n$ 
\cite{Manton,Liang}.) We may denote a solution specified by 
eq.(\ref{radius}) with an integer $n$ by $\phi_n (y)$.
We note that as $k$ runs from zero to one the right hand side of
eq.(\ref{radius}) increases monotonically from $R^*_n
\equiv (n-\half)/\mu$ to infinity. Thus, $\phi_n(y)$ is a solution
only when $R \geq R^*_n$.
\par
For $0 < R \leq R^*_1$, there exists only one solution to the field
equation (\ref{fieldeq}), i.e. the trivial solution $\phi = 0$. 
Thus,
the vacuum configuration is given by the trivial solution,
and hence 
the translational invariance is unbroken for $0 < R \leq R^*_1$. 
For
$R^*_1 < R \leq R^*_2$, there exist two solutions to eq.(\ref{fieldeq}),
i.e. the trivial one and $\phi_1(y)$. It follows from eq.(\ref{energy4})
that the trivial solution $\phi = 0$ is no longer the vacuum 
configuration. 
Since $\phi_1(y)$ depends on $y$, the translational
invariance for the $S^1$-direction is spontaneously broken. 
For $R^*_n < R \leq R^*_{n+1}$, there exist $n+1$ solutions to
eq.(\ref{fieldeq}), i.e. the trivial one and $\phi_m(y)$ for $m =
1,2, \cdots ,n.$ 
Since  ${\cal E}[\phi_m,R] < {\cal E}[0,R]$ for
every $m$, the trivial solution is no longer the vacuum configuration,
and hence the translational invariance for the $S^1$-direction
is spontaneously broken
for $R^*_n < R \leq R^*_{n+1}$. 
Therefore, we have found that for 
$0< R \leq R^*_1$ the translational invariance 
for the $S^1$-direction is unbroken,
while for $R > R^*_1$ it is broken spontaneously with the $Z_2$
symmetry breaking.
\par
Let us next discuss a problem which solution $\phi_n (y)$ is the
true vacuum configuration, i.e. which solution minimizes the functional
${\cal E}[\phi,R]$. For $R > R^*_n$, $\phi_n (y)$ becomes a solution
for which we obtain a rather complicated expression
\BQ
{\cal E}[\phi_n,R] = \frac{(2n-1)\mu^3}{3\lambda(1+k^2)^{3/2}}
\left\{-(1-k^2)(5+3k^2)K(k)+8(1+k^2)E(k) \right\},
\label{energy5}
\EQ
where $E(k)$ is the complete elliptic function of the second kind.
Although we could directly compare ${\cal E}[\phi_n,R]$ for 
$n=1,2,3,\cdots,$ we shall here take another approach to solve the 
problem. It is not difficult to show
\BQ
\frac{d{\cal E}[\phi_n, R]}{dR}
= \frac{\pi \mu^4}{\lambda}\left(\frac{1-k^2}{1+k^2} 
\right )^2 \geq 0,
\label{differ}
\EQ
which implies that ${\cal E}[\phi_n, R]$ is a monotonically increasing
function of $R$. At $R = R^*_n$ $(k=0)$ and $\infty$ $(k=1)$,
${\cal E}[\phi_n, R]$ takes the values
\BQA
{\cal E}[\phi_n, R^*_n] &=& (2n-1)\frac{\pi \mu^3}{2\lambda}, \NN
{\cal E}[\phi_n, \infty] &=& (2n-1)\frac{4\sqrt{2} \mu^3}{3\lambda},
\label{energy6}
\EQA
respectively.
It follows from eqs.(\ref{differ}) and (\ref{energy6}) that
\BQ
(2n-1)\frac{\pi \mu^3}{2\lambda} \leq
{\cal E}[\phi_n, R] < (2n-1)\frac{4\sqrt{2} \mu^3}{3\lambda}
\label{ineq}
\EQ
and especially
\BQ
{\cal E}[\phi_1, R] < \frac{4\sqrt{2} \mu^3}{3\lambda}.
\label{ineq2}
\EQ
The above observations will be enough to show that for $R \geq R^*_n$
($n \geq 2$)
\BQ
{\cal E}[\phi_1, R] < {\cal E}[\phi_n, R] < {\cal E}[0, R].
\label{ineq3}
\EQ
Therefore, we have found the vacuum expectation value of $\phi$
to be
\BQ
\langle \phi(x^{\mu}, y) \rangle
= {0  \quad \qquad {\rm for} \quad R \leq R^*_1
\atopwithdelims\{. \phi_1(y) \qquad {\rm for }\quad R > R^*_1.}
\label{vev}
\EQ
\par
It may be instructive to reanalyze the model from 
a viewpoint of the Fourier expansion.
It follows from the boundary condition (\ref{bc2})
that $\phi(y)$ may be expanded in the Fourier-series
for the $S^1$-direction as 
\[
\phi(y) = \frac{1}{\sqrt{\pi R}}\sum^{\infty}_{l=1}
\left\{a^{(2l-1)}\cos\left((2l-1)\frac{y}{2R}\right)+
b^{(2l-1)}\sin\left((2l-1)\frac{y}{2R}\right)\right\},  
\]
or equivalently,
\BQ
\phi(y) = \frac{1}{\sqrt{2\pi R}}\sum_{l \in Z}\varphi^{(2l-1)}
e^{i(2l-1)\frac{y}{2R}}
\label{mode2}
\EQ
with $\varphi^{(2l-1)}$ = $\frac{1}{\sqrt{2}}(a^{(2l-1)}-ib^{(2l-1)})$
= $\varphi^{(-2l+1)*}$. A key observation is that a constant zero mode
is excluded in the above expansion due to the nontrivial boundary
condition. 
Inserting eq.(\ref{mode2}) into eq.(\ref{energy}), we have,
up to the quadratic terms with respect to $\varphi^{(2l-1)}$,
\BQ
{\cal E}_0[\varphi, R] = \sum^{\infty}_{l=1} m^2_{(2l-1)}
|\varphi^{(2l-1)}|^2,
\label{energy7}
\EQ
where
\BQ
m^2_{(2l-1)} \equiv -\mu^2 + \left(\frac{2l-1}{2R}\right)^2.
\label{mass}
\EQ
The second term in eq.(\ref{mass}) is the Kaluza-Klein mass, which
comes from the \lq \lq kinetic'' term $\half \left(\del_y \phi(y)
\right)^2$, and which gives a positive contribution to the squared 
mass term. We can easily see that for $R \leq R^*_1$ all the squared
masses $m^2_{(2l-1)}$ are positive semi-definite because of the
induced Kaluza-Klein masses $\left(\frac{2l-1}{2R}\right)^2$.
On the other hand, for $R > R^*_1$ it seems that negative squared
masses appear. 
This is a signal of a phase transition and is
consistent with the results obtained before.
\par
It may be interesting to point out that the translational invariance
for the $S^1$-direction
can be reinterpreted as a global $U(1)$ symmetry,
which is in fact possessed by the theory after the compactification.
To see this,
we note that an infinitesimal translation $y \rightarrow y+a$ in
eq.(\ref{mode2}) can equivalently be realized, in terms of the
Fourier modes, by the following transformation:
\BQ
\varphi^{(2l-1)} \longrightarrow 
e^{i(\frac{2l-1}{2R})a}\varphi^{(2l-1)},
\label{transf}
\EQ
from which we may assign a $U(1)$ charge $\frac{2l-1}{2R}$ to
$\varphi^{(2l-1)}$. Thus, spontaneous breakdown of the translational 
invariance for the $S^1$-direction may be interpreted as that of the $U(1)$
symmetry. One might then ask about a Nambu-Goldstone mode associated
with spontaneous breakdown of 
the translational invariance or the $U(1)$
symmetry. It turns out that, to answer the question, the following
mode expansion is more suitable than the Fourier mode expansion
for $R > R^*_1$ \cite{Manton,Liang}:
\BQ
\phi(y)=\sum^{\infty}_{l=1}\left\{{a'}^{(2l-1)}Ec^{2l-1}(\omega y,k)
+{b'}^{(2l-1)}Es^{2l-1}(\omega y,k)\right\},
\label{lame}
\EQ
where $Ec^{2l-1}(u, k)$ and $Es^{2l-1}(u, k)$ are eigenfunctions
of the so-called Lam\'e equation with $N = 2$ \cite{Bate}
\BQ
\left[-\frac{d^2}{du^2}+N(N+1)k^2{\rm sn}^2(u,k)\right]\Psi(u,k)
= \Omega(k)\Psi(u,k),
\label{lameeq}
\EQ
with the boundary condition
\BQ
\Psi(u+2K(k),k) = -\Psi(u,k).
\label{bc3}
\EQ
The $Ec^{2l-1}(u, k)$ and $Es^{2l-1}(u, k)$ may further be 
supplemented by the following conditions
\footnote{The eigenfunctions $Ec$ and $Es$ are differently
defined from those in ref.\cite{Bate}.}:
\BQA
Ec^{2l-1}(-u, k) &=& +Ec^{2l-1}(u, k),  \NN
Es^{2l-1}(-u, k) &=& -Es^{2l-1}(u, k),
\EQA
and
\BQA
Ec^{2l-1}(u, k) &\longrightarrow& 
\frac{1}{\sqrt{\pi R}}\cos\left((2l-1)u \right) 
\qquad {\rm as} \quad k \rightarrow 0, \NN
Es^{2l-1}(u, k) &\longrightarrow& 
\frac{1}{\sqrt{\pi R}}\sin\left((2l-1)u \right) 
\qquad {\rm as} \quad k \rightarrow 0.
\label{limit}
\EQA
In the expansion (\ref{lame}), ${a'}^{(2l-1)}$ and ${b'}^{(2l-1)}$
correspond to normal modes around the background $\phi=\phi_1(y)$.
If we set $\Omega(k) = (1+k^2)(1+\frac{m^2}{\mu^2})$, $m^2$ may
correspond to a squared mass in ($D-1$)-dimensional Minkowski
spacetime. 
The lowest five eigenvalues and eigenfunctions for the
Lam\'e equation with $N=2$ are exactly known,
and the eigenfunctions are given by so-called 
Lam\'e polynomials
\cite{Bate}. Only two of them satisfy the desired boundary condition
(\ref{bc3}), and are given by 
$Ec^{1}(u,k) \propto {\rm cn}(u,k){\rm dn}(u,k)$ and
$Es^{1}(u,k) \propto {\rm sn}(u,k){\rm dn}(u,k)$ with $m^2 = 0$ and 
$\left(\frac{3k^2}{1+k^2}\right)\mu^2$, respectively. Noting that
$Ec^{1}(\omega y, k) \propto \frac{d\phi_1(y)}{dy}$, we know that the
mode  ${a'}^{(1)}$ is really the massless Nambu-Goldstone mode associated
with spontaneous breakdown of 
the translational invariance or the 
$U(1)$ symmetry.
\par
We have shown that the translational invariance for the 
$S^1$-direction is 
spontaneously broken 
in the model (\ref{action}) with the boundary
condition (\ref{bc2})
when the radius $R$ becomes larger than a critical
radius $R^*_1$. Our mechanism to break the translational invariance is not 
specific to this model at all. 
Let us briefly discuss a general
strategy to construct models in which the translational invariance of
compactified spaces can be broken spontaneously. Suppose that
some of space dimensions are compactified on a manifold with the translational
invariance. Let $V(\phi_i)$ be a scalar potential. Our mechanism
may require $V(\phi_i)$ to satisfy the following two conditions:

\begin{enumerate}
\item The origin $\phi_j = 0$ is not the minimum of the potential
        $V(\phi_i)$.
\item
Let $\bar{\phi_j}$ be a configuration which minimizes $V(\phi_i)$.
Then, some of $\phi_j$ with $\bar{\phi_j} \neq 0$ have to be non-singlets
for some global symmetries of the theory.
\end{enumerate}
\par
A key ingredient of our mechanism is to impose nontrivial boundary
conditions on non-singlet fields $\phi_j$ with $\bar{\phi_j} \neq 0$,
which have to be consistent with global symmetries of the theory.
We would have a variety of models since we have a wide choice of
potentials, compactified spaces and boundary conditions. A general
feature of our models will be that the translational invariance of 
compactified spaces is expected to be unbroken when scales of
the compactified spaces are sufficiently small and to be broken
spontaneously with some global symmetries when the scales become
large enough.
\par
Finally, we would like to make some comments on vacuum structures
of our models and on an application to supersymmetric field theories.
In the limit of $R \rightarrow \infty (k \rightarrow 1)$, $\phi(y)$
in eq.(\ref{sol}) will reduce to 
$\sqrt{\frac{2}{\lambda}}\mu \tanh\left(\frac{\mu}{\sqrt{2}}
(y-y_0)\right)$. This is nothing but a (static) single kink solution
in $D = 2$ dimensions \cite{Raja}. So, the model considered here
may be regarded as a real $\phi^4$ model on a single kink background
sitting on a line in the limit of $R \rightarrow \infty$. This
observation suggests that models based on our breaking mechanism
might be regarded as quantum field theories on (topologically)
nontrivial backgrounds {\it in a broken phase} of the translational
invariance. 
The second comment is that vacuum structures of our
models are expected to be quite nontrivial, in general. To see this,
let us consider a simple extension of the model (\ref{action}) by
replacing the real field $\phi$ by a complex one $\Phi$ with a 
$U(1)$ symmetry and also the boundary condition (\ref{bc2}) by
$\Phi(y+2\pi R) = e^{i2\pi \alpha}\Phi(y)$. One might expect that
the vacuum structure is similar to the original real $\phi^4$ model.
This is not, however, the case. 
In fact, as studied in ref.\cite{Tomaras}, 
this model admits a rich set of solutions to the field
equation for $\Phi(y)$, and the vacuum configuration is quite
different from that of the real $\phi^4$ model
\footnote{Since in ref.\cite{Tomaras} the field equation has been
solved with the periodic boundary condition, we should reanalyze
the field equation with a proper boundary condition.}. 
The final
comment is that it would be worth applying our mechanism
to supersymmetric field theories. In ref.\cite{Sakamoto}, it has
been shown that our mechanism can be used to break supersymmetry
spontaneously and that this SUSY breaking mechanism is new, that is,
it is different from the O'Raifeartaigh \cite{OR} and Fayet-Iliopoulos
\cite{FI} mechanisms. It would be of great importance to use this new 
SUSY breaking mechanism to construct phenomenologically interesting
supersymmetric models.
\vskip0.3truein
\centerline{{\it ACKNOWLEDGMENTS}}
We would like to thank to H. Hatanaka and C. S. Lim
for useful discussions.
K.T. would like to thank the I.N.F.N, Sezione di Pisa for hospitality.
\newpage

\end{document}